\documentstyle[sprocl,epsf]{article}

\newcommand{\Ibdz}{\int\,{\rm d^2}b\;}

\newcommand{\Tr}{\makebox{ Tr }}

\newcommand{\GeV}{\makebox{ GeV}}

\newcommand{\fm}{\makebox{ fm}}

\newcommand{\beq}{\begin{equation}}

\newcommand{\enq}{\end{equation}}

\newcommand{\beqa}{\begin{eqnarray}}

\newcommand{\enqa}{\end{eqnarray}}

\newcommand{\nn}{\nonumber}

\newcommand{\lbq}[1]{\label{#1} \enq}

\newcommand{\lbqa}[1]{\label{#1} \enqa}

\newcommand{\befi}[1]{\begin{figure}[ht] \leavevmode \centering \epsffile{#1.eps}}

\newcommand{\lbfi}[1]{\label{#1} \end{figure}}

\newcommand{\eq}[1]{eq.(\ref{#1})}

\newcommand{\fig}[1]{fig.(\ref{#1})}

\newcommand{\ct}{\cite}

\newcommand{\lbcap}[3]{\begin{minipage}{#1}\caption{\small #2}\label{#3}\end{minipage}\end{figure}}

\newcommand{\pa}{\partial}

\newcommand{\cP}{\mbox{$\cal P$}}

\newcommand{\cS}{\mbox{$\cal S$}}

\newcommand{\bA}{\mbox{\bf A}}

\newcommand{\bV}{\mbox{\bf V}}

\newcommand{\bW}{\mbox{\bf W}}

\newcommand{\bo}{\mbox{\bf 1}}

\newcommand{\al}{\alpha}

\newcommand{\be}{\beta}

\newcommand{\de}{\delta}

\newcommand{\ep}{\epsilon}

\newcommand{\la}{\lambda}

\newcommand{\rh}{\rho}

\newcommand{\si}{\sigma}

\newcommand{\De}{\Delta}

\newcommand{\Ps}{\Psi}

\begin{document}
\title{\Large \bf Hadron structure and odderon exchange}
\author{\large Michael Rueter}
\address{\vspace*{2mm}Institut f\"ur theoretische Physik\\
Universit\"at Heidelberg\\
Philosophenweg 16, D-69120 Heidelberg, FRG\\
e-mail: M.Rueter@thphys.uni-heidelberg.de\\[2mm]
\rm Talk given at the {\it Diquarks III} workshop, Torino, October 1996\\[2mm]
\rm supported by the Deutsche Forschungsgemeinschaft
}
\maketitle\abstracts{We calculate the $C$=$P$=--1 contribution to high-energy scattering of hadrons in the framework of the model of the stochastic vacuum. In models, where the pomeron is generated by a two gluon exchange, this odderon contribution comes from a three gluon exchange and is much too large as compared to experimental data for $pp$- and $p\bar{p}$-scattering. In our model, where we have no quark-additivity, the hadron structure is very important. It is shown that a natural suppression of the odderon contribution is given by a diquark-structure of the nucleon.}
In this note we summarize the results presented in a recent publication\cite{Rueter:1996}. Some parameters of the used model are changed but the calculated physical quantities are rather insensitive to these changes.
\section{Introduction}
The possibility that the real part of the scattering amplitude increases with energy as fast as the imaginary part was first considered by Lukaszuk and Nicolescu\ct{Lukaszuk:1973}. Such a behavior would mean that a trajectory of a pole which is odd under $C$ and $P$ has an intercept near one. This trajectory has been called odderon. One consequence of such an odderon would be that the ratio of the real to imaginary part of the forward scattering amplitude is different for particle-particle and particle-antiparticle-scattering even at asymptotic energies. For further reference we shall use the conventional abbreviation $\De \rh$ for that difference: 
\beq
\De \rh(s) = \rh^{\bar p p}(s) - \rh^{p p}(s)=\frac{{\rm Re}\left[ T^{\bar p p}(s,0)\right]}{{\rm Im}\left[  T^{\bar p p}(s,0)\right]}-\frac{{\rm Re}\left[ T^{ p p}(s,0)\right]}{{\rm Im}\left[ T^{ p p}(s,0)\right]}
\lbq{rhodef}
Interest in the odderon rose again when the UA4 collaboration\ct{Bernard:1987} reported a value for $\rh^{\bar p p}$ at $\sqrt{s} = 546$ GeV which was much larger than the one extrapolated by means of dispersion relations for proton-proton-scattering and thus seemed to indicate a large value for $|\De \rh|$.\\
The new results of the UA4/2 collaboration\ct{Augier:1993} obtained however a value $\rh^{\bar p p}(\sqrt{s} = 541 \GeV)= 0.135 \pm 0.015$ which is very well compatible with $\De \rh = 0$ at that energy\cite{Bourrely:1984}$\mbox{}^-$\nocite{Donnachie:1984}\nocite{Bourrely:1987}\nocite{Gauron:1988}\nocite{Kroll:1989}\cite{Covolan:1993} and at any rate leaves no room for a large value of that quantity. The very successful description of high-energy data by the Donnachie-Landshoff pomeron\ct{Donnachie:1992} also yields $\De \rh \approx 0$.\\
As far as the contribution of three {\em non-perturbative} gluons is concerned there is no reason for a strong suppression of the three gluon versus the two gluon exchange. In an Abelian model for non-perturbative gluon exchange\ct{Landshoff:1987} Donnachie and Landshoff\ct{Donnachie:1991} have found that the lowest order effective odderon coupling, i.e.~the coupling of three non-perturbative gluons, is suppressed by a factor of two with respect to the effective pomeron coupling. Though it is very gratifying that in a non-perturbative model the three gluon coupling is smaller than the two gluon coupling (naive expectation goes in the opposite direction), this coupling still leads to a value of $|\De \rh | \approx .5$ which is far from consistency with the analysis of the data. In the Abelian model of Landshoff and Nachtmann, where quark additivity is a consequence of the model, the $\rh$-parameter for hadron-(anti)hadron-scattering is just the one for quark-(anti)quark-scattering.\\
In a series of papers\ct{Dosch:1994} a non-Abelian model of high-energy scattering was presented which gives a good description of the experimental data and relates parameters of high-energy scattering to those of hadron spectroscopy. In this model, called the model of the stochastic vacuum\ct{Dosch:1987}$\mbox{}^,$\ct{Dosch:1988} (MSV), the non-perturbative gluonic contributions to the QCD-pathintegral are approximated by a Gaussian stochastic process with a gluonic correlation length $a$. One of the most characteristic features of this model is that the same mechanism which leads to confinement introduces a kind of string-string interaction in high-energy scattering and leads to a marked increase of the total cross section as a function of the hadron size even if the latter is large as compared to the correlation length $a$. Quark-additivity does not hold in that approach. The different total cross sections for pion-nucleon-, kaon-nucleon- and nucleon-(anti)nucleon-scattering are correctly reproduced due to the different radii of the hadrons. In this note we review the evaluation of the leading $C$=$P$=--1 contribution of that model. We show that this contribution (and therefore also $\De \rh$) depends crucially on the structure of the nucleon; we especially discuss the dependence of $\De \rh$ on the radius of a diquark if two quarks are clustered.
\section{Diffractive high-energy scattering of hadrons in the MSV}
To calculate the diffractive scattering amplitude in the high-energy limit (\mbox{$s\to \infty$}) one uses an eikonal approximation to separate the large energy scale $s$ from the small momentum transfer scale $t$ in a fixed gluon background field\cite{Nachtmann:1991}. The resulting quark-quark-scattering amplitude (with helicity $\la_i$ and color $C_i$) is very similar to the quantum-mechanical scattering in an external potential:
\beqa
T(s,t)&=&2\,i\,s\,\de_{\la_3\la_1}\de_{\la_4\la_2}\, \Ibdz e^{-i\vec{q}\,\vec{b}} \hat{J}\nn\\
\mbox{where }\hat{J}&=&< \left( \left[ \bV [P_1]-\bo
\right]_{C_3C_1}\left[ \bV [P_2]-\bo\right]_{C_4C_2}\right) >\nn\\
\mbox{and }\bV [P_i]&=& \cP\, \exp\left(-ig\int_{P_{i}}\bA_\mu(z) {\rm d}z^\mu\right).
\lbqa{Jquark}
The quark picks up the phase $\bV [P_i]$ by interacting with the background field on the light-like path $P_i$. The momentum transfer is given by $t=-\vec{q}^2$ and $\vec{b}$ is the impact-parameter. The scattering amplitude is proportional to $s$ and we have helicity conservation. The problem is to calculate the expectation value $<\ldots >$ with respect to the background field. We do this by using the MSV. In this non-Abelian model it is crucial to respect gauge invariance. We go from quark-quark- to hadron-hadron-scattering by constructing as usual gauge invariant hadrons by connecting the constituents with Schwinger-strings. For mesons we end up with the ``scattering'' of Wilson-loops\ct{Dosch:1994} (see \fig{2loops}). The meson-scattering amplitude is obtained by averaging the loop-scattering with Gaussian transversal wave function with extension parameter $S$:
\beqa
\bW [\pa S_i]&=& \cP\, \exp\left(-ig\oint_{\pa S_{i}}\bA_\mu(z) {\rm d}z^\mu\right)\nn\\
\tilde{J}(\vec{b},\vec{R}_1,\vec{R}_2)&=&-<\frac{1}{N_C}\Tr \left[ \bW [\pa S_1]-{\bf 1}\right]\cdot\frac{1}{N_C}\Tr \left[ \bW [\pa S_2]-{\bf 1}\right]>\nn\\
\hat{J}(\vec{b},S_1,S_2)&=&\int\,{\rm d^2}\vec{R}_1\int\,{\rm d^2}\vec{R}_2\,\tilde{J}(\vec{b},\vec{R}_1,\vec{R}_2) \,|\Ps(\vec{R}_1,S_1)|^2\,|\Ps(\vec{R}_2,S_2)|^2\nn\\
\Ps(\vec{R}_i,S_i)&=& \sqrt{\frac{1}{2\pi}}\frac{1}{S_i}e^{-\frac{|\vec{R}_i|^2}{4S_i^2}}.
\lbqa{Jmeson}
\epsfxsize6cm
\befi{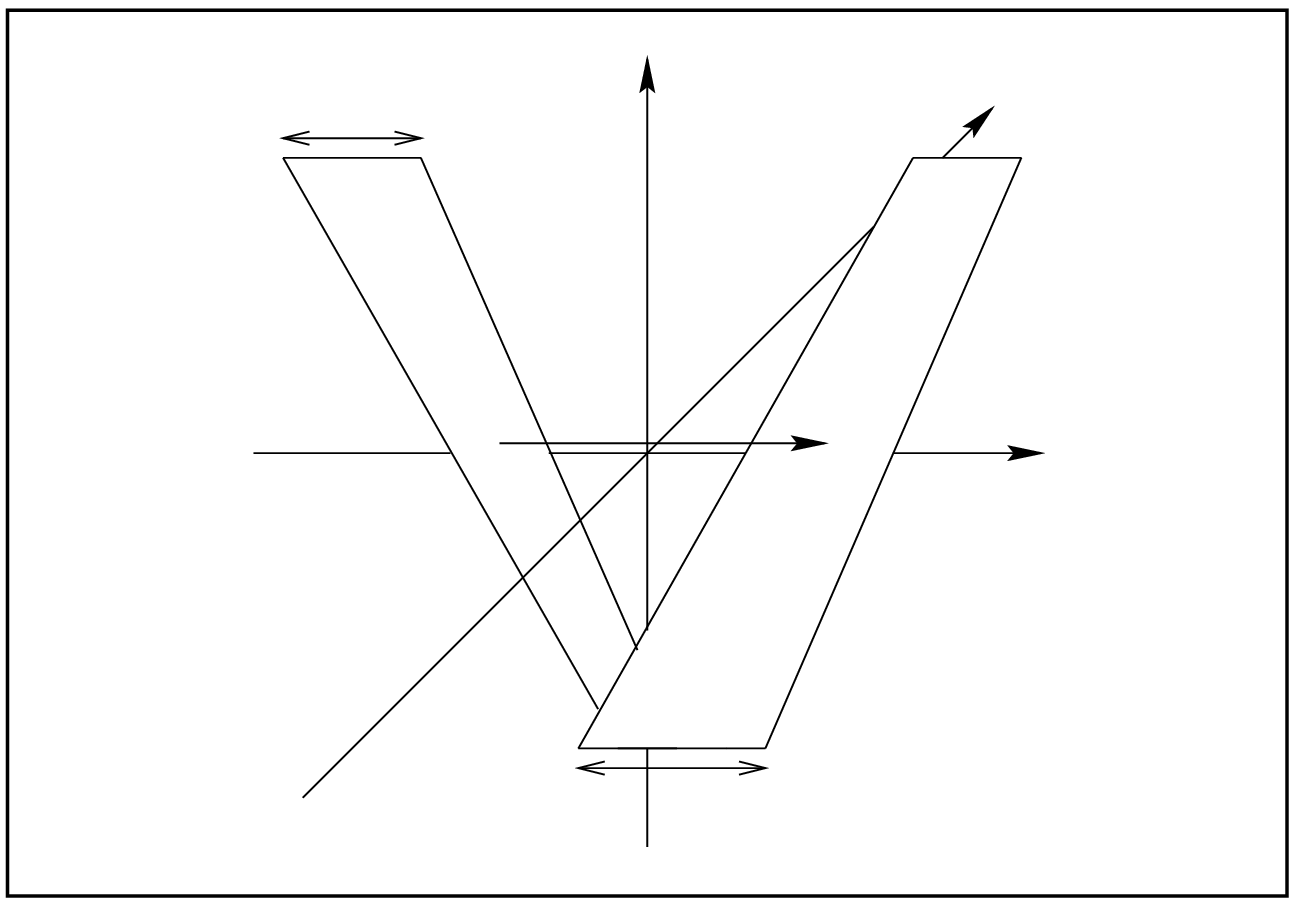}
\unitlength.6cm
\begin{picture}(0,0)
\put(-8.9,4.5){loop 1}
\put(-2,4.5){loop 2}
\put(-2.2,3){$\vec{x}$}
\put(-4.9,6.2){$x^0$}
\put(-2.2,6){$x^3$}
\put(-7.5,6.1){$\vec{R}_1$}
\put(-4.9,.4){$\vec{R}_2$}
\put(-5.4,3.7){$\vec{b}$}
\end{picture}
\caption{Two loops with transversal extension $\vec{R}_1$ and $\vec{R}_2$ and light-like sides. Loop 1 ($\pa S_{1}$) describes a colorless $q\bar{q}$-pair running in negative 3-direction and loop 2 in positive 3-direction. The impact parameter $\vec{b}$ is chosen to be purely transverse and the light-like sides are considered to by infinitely long.}
\lbfi{2loops}
\\In order to describe baryon-baryon-scattering, one has to start from three loops (without traces) with one common side as shown in \fig{3loops}.
\epsfxsize6cm
\befi{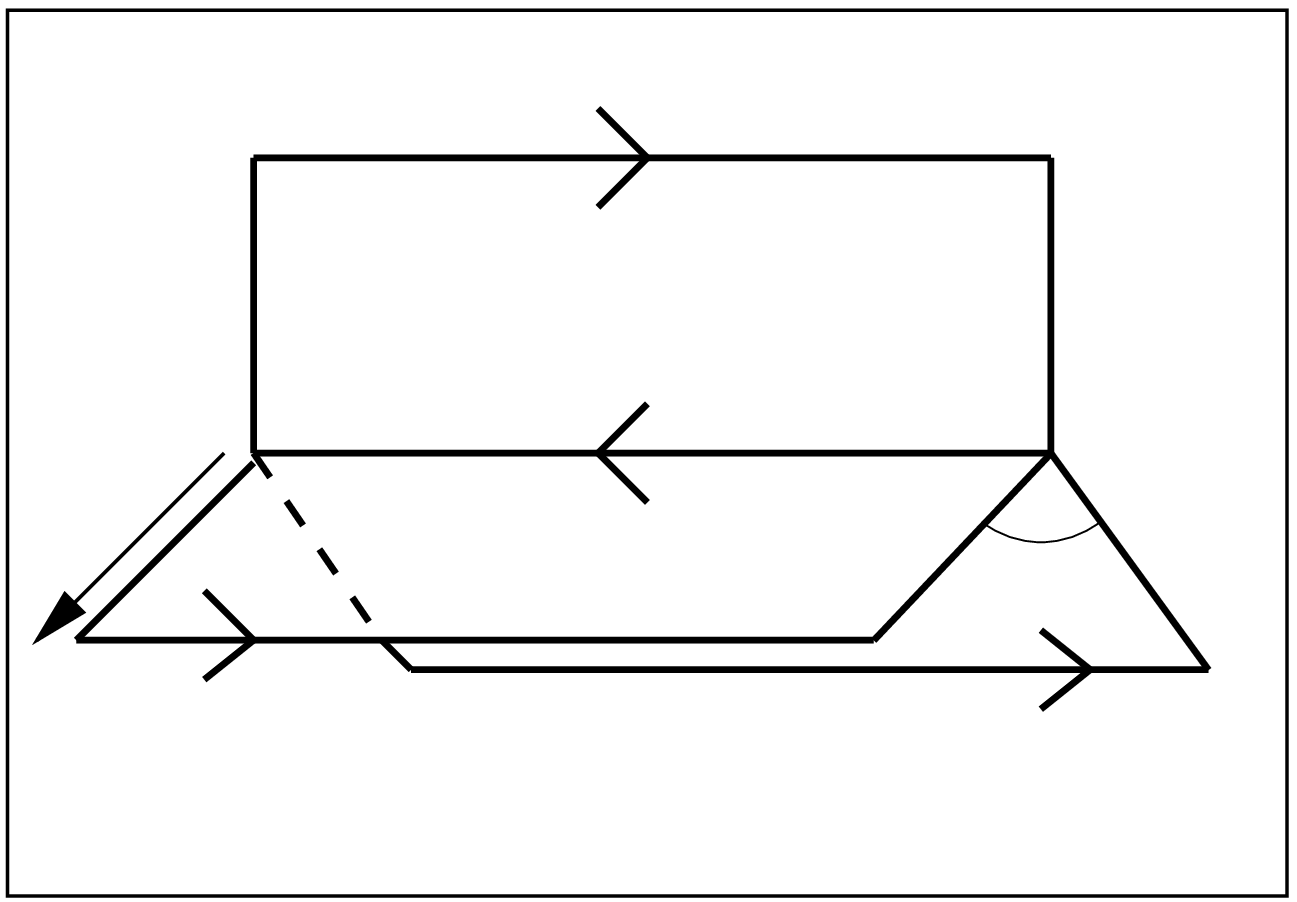}
\unitlength.813cm
\begin{picture}(0,0)
\put(-1.6,2.2){$\al$}
\put(-1.6,.7){$\pa \cS_{i2}$}
\put(-6.6,.9){$\pa \cS_{i1}$}
\put(-3.6,4.5){$\pa \cS_{i3}$}
\put(-7.1,2.4){$\vec{R}_i$}
\end{picture}
\caption{The colorless $qqq$-objekt is constructed out of 3 loops with one common line which transforms like a color singlet. Here $\pa \cS_{ij}$ denotes the loop corresponding to quark j of \mbox{baryon i}. By varying the angle $\al$ we can consider different geometries of the baryon. With $\vec{R}_i$ we denote the radius. A baryon is obtained by averaging with a transversal wave function.}
\lbfi{3loops}
\\The reduced scattering amplitude for a colorless $qqq$-objekt is given by:
\beqa
\tilde{J}(\vec{b},\vec{R}_1,\vec{R}_2)&=&-<B_1\cdot B_2>\nn\\
{\rm with }\;B_i&=&\frac{1}{6}\ep_{abc}\ep_{a'b'c'}\left\{ \bW_{a'a}[S_{i1}]\bW_{b'b}[S_{i2}]\bW_{c'c}[S_{i3}]-\de_{a'a}\de_{b'b}\de_{c'c}\right\}\nn\\
\bW_{a'a}[S_{ij}]&=& \left[\cP\, e^{-ig\oint_{\pa S_{ij}}\bA_\mu (z)\,{\rm d}z^\mu}\right]_{a'a}.
\lbqa{Jbaryon}
Baryon-baryon-scattering is obtained by averaging with a transversal wave function:
\beq
\Ps(\vec{R}_i,S_i)= \sqrt{\frac{2}{\pi}}\frac{1}{S_i}e^{-\frac{|\vec{R}_i|^2}{S_i^2}}.
\lbq{Gausswfbaryon}
We consider two classes of baryon configurations. In the first case the distances from the common line of all three loops are equal and we vary the angle $\al$ between two loops (see \fig{3loops}). If this angle tends to zero the two quarks together form a point-like diquark (i.e.~an object transforming under the \mbox{$\bar{3}$-representation} of $SU(3)$). The other case we consider is a linear structure of the nucleon (see \fig{ort}).
\epsfxsize6cm
\befi{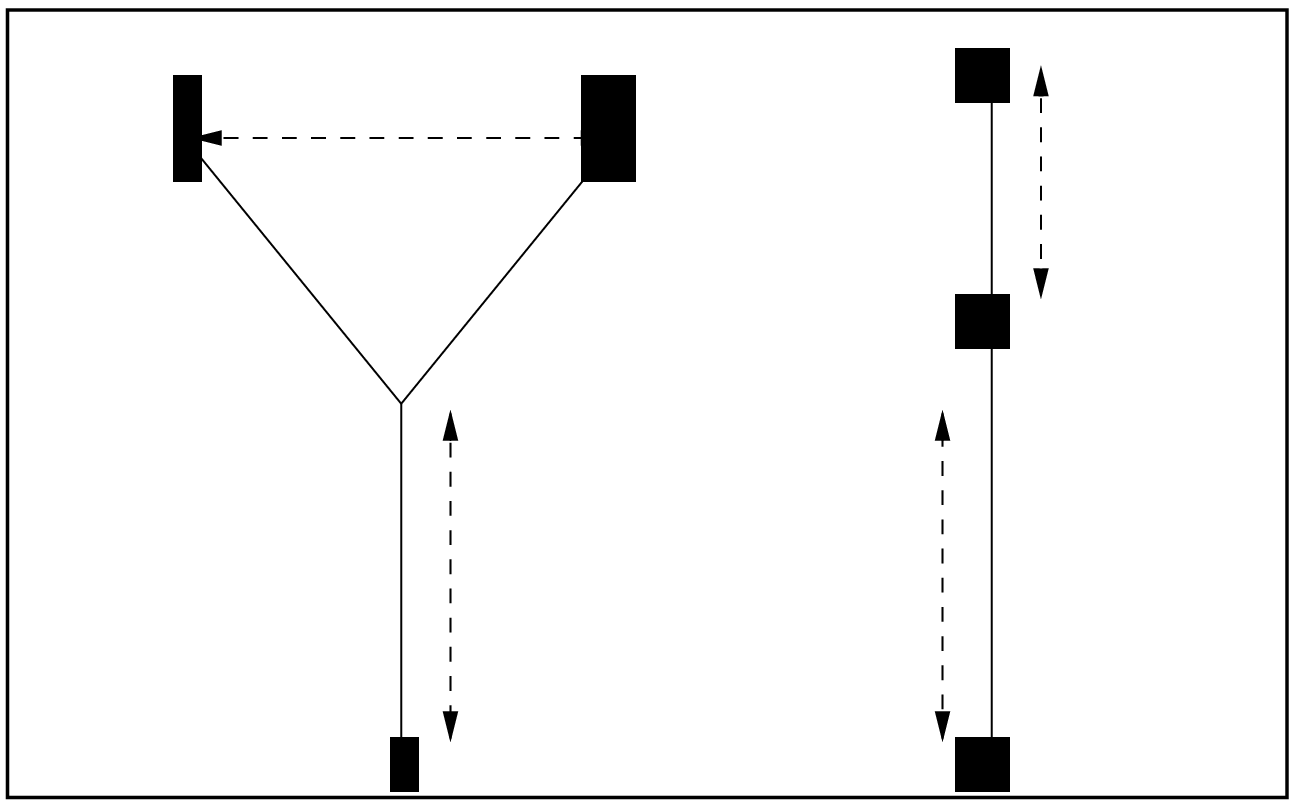}
\unitlength.462cm
\begin{picture}(0,0)
\put(-8.2,2.2){$R$}
\put(-4.4,2.2){$R$}
\put(-9.2,7){$r_\perp$}
\put(-2.4,6.2){$r_\|$}
\put(-9.2,4.5){$\al$}
\end{picture}
\caption{In this figure we show the star-like and linear geometry for the baryon. For small quark distances $r_\perp$ and $r_\|$ we have a diquark configuration.}
\lbfi{ort}
\\In the limit of the angle $\al$ (see \fig{3loops}) going to zero, the baryon can effectively be treated like a meson, the resulting diquark playing the role of the antiquark. If $\al \rightarrow 0$ the loop $\pa \cS_{i2}$ goes over into $\pa \cS_{i1}$ and we have
\beq
B_i(\al = 0)=\frac{1}{6}\ep_{abc}\ep_{a'b'c'}\left\{ W_{a'a}[\cS_{i1}]W_{b'b}[\cS_{i1}]W_{c'c}[\cS_{i3}]-\de_{a'a}\de_{b'b}\de_{c'c}\right\}.
\lbq{3a}
We use the following identity for an arbitrary $SU(3)$-matrix $U$
\[
\ep_{a'b'c'}U_{a'a}U_{b'b}U_{c'c}=\ep_{abc}
\]
from which we obtain
\beq
\ep_{a'b'c'}W_{a'a}[\cS_{i1}]W_{b'b}[\cS_{i1}]=\ep_{abh}W^{-1}_{hc'}[\cS_{i1}]= \ep_{abh}W_{hc'}[\cS_{i1}^{-1}]
\lbq{3b}
where $\pa \cS_{i1}^{-1}$ is the Wilson-loop oriented in opposite direction. Inserting \eq{3b} in \eq{3a} we obtain
\beqa
B_i(\al = 0)&=&\frac{1}{3}\de_{ab}\left\{ W_{ac}[\cS_{i1}^{-1}]W_{cb}[\cS_{i3}]-\de_{ab}\right\}=\frac{1}{3}\de_{ab}\left\{ W_{ab}[\hat{\cS}_{i12}]-\de_{ab}\right\}\nn\\
&=&\frac{1}{3}\Tr \left\{\bW[\hat{\cS}_{i12}]-\bo\right\}
\lbqa{3c}
where $\pa\hat{\cS}_{i12}$ is the union of $\pa \cS_{i1}^{-1}$ and $\pa \cS_{i3}$. This is exactly the contribution of a quark traveling along line 3 and an antiquark traveling along line 1=2. The spin contribution in the high energy limit of a quark and antiquark is equal, namely $2s\de_{\la\la'}$, where $\la$ is the helicity in the initial and final state respectively. Thus in the limit $\al \rightarrow 0$ the baryon can be treated effectively as a meson, the point like diquark traveling along line 1=2 replacing the antiquark of the meson.
\section{Results for the total cross-section and the slope}
To calculate the vacuum expectation values in \eq{Jmeson} and \eq{Jbaryon} we use the model of the stochastic vacuum. One approximates the averaging by a Gaussian stochastic process for the parallel transported fieldstrengths. This process is mainly characterized by the correlation length $a$ and the usual gluon condensate $<g^2FF>$. For details we have to refer to the literature\ct{Dosch:1987}$\mbox{}^,$\ct{Dosch:1988}$\mbox{}^,$\cite{Dosch:1994}. The parameters are fixed by comparison with experimental data for high energy scattering and with hadron spectroscopy\cite{Dosch:1994}$\mbox{}^,$\cite{Rueter:1996}. A new analysis for the different baryon geometries yields\cite{Rueter:1997}:
\begin{table}[ht]
\centerline{
\begin{tabular}{|c||c|c|c|}\hline
&diquark&star-like&linear\\ \hline\hline
$<g^2FF>$&3.0$\GeV ^4$&3.0$\GeV ^4$&3.1$\GeV^4$\\ \hline
$a$&0.31 fm&0.30 fm&0.33 fm\\ \hline
\end{tabular}}
\caption{The parameter sets}
\label{parameter}
\end{table}
\\It turns out that for the total cross section the different geometries give almost the same results. For simplicity we consider here only the diquark structure. By expanding the exponentials in \eq{Jmeson} up to second order we find the leading contribution to the scattering amplitude. This contribution is purely imaginary and even under charge parity. With the help of the optical theorem we find a total cross section which depends only on the extension parameter $S$ and is the same for hadron-hadron- and hadron-antihadron-scattering. The numerically obtained total cross section can be parameterized by
\[
\si^{\rm tot} = <g^2FF>^2 a^{10}\, \al\,\left( \frac{S}{a}\right)^{\displaystyle \be}
\]
where
$\al=5.20*10^{-3}$ and $\be =3.03$. The result is shown in \fig{sigtot}.
\epsfxsize6cm
\befi{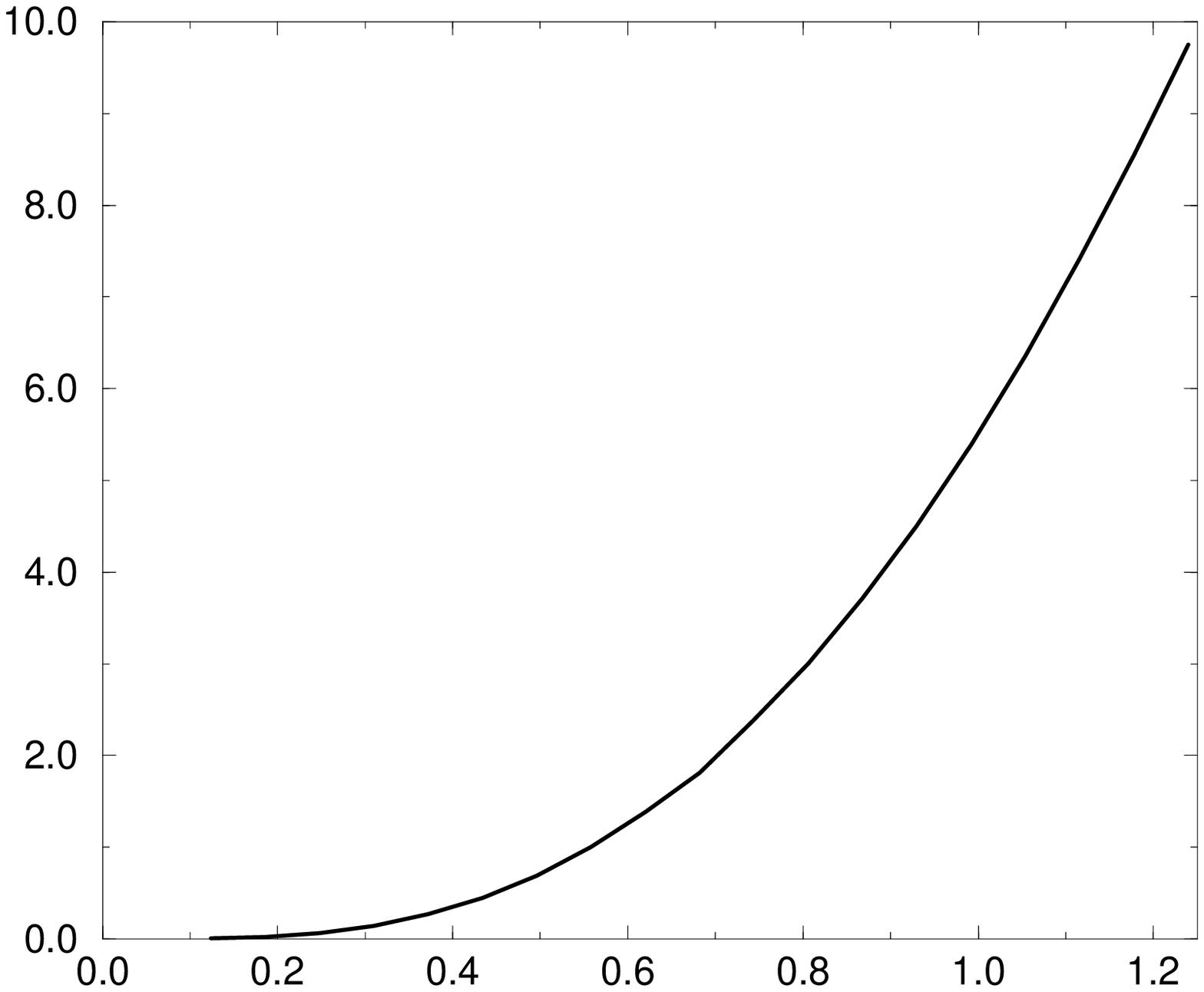}
\unitlength.9cm
\begin{picture}(0,0)
\put(-1.5,1.){$S[\fm ]$}
\put(-5.8,4.7){$\si^{\rm tot}[\fm^2]$}
\end{picture}
\caption{The total cross section of hadron-hadron-scattering with diquark geometry as a function of the extension parameter $S$.}
\label{sigtot}
\end{figure}
We also calculated the slope B,
\beq
B=\frac{\rm d}{{\rm d}t}\ln\left( \frac{{\rm d}\si^{\rm el}}{{\rm d}t} \right)|_{t=0},
\lbq{slopedef}
which in our model also only depends on the extension parameter $S$. It was shown\cite{Dosch:1994}$\mbox{}^,$\cite{Rueter:1997} that the model reproduces the experimental data for $p$-$p$-, $p$-$\bar{p}$-, $p$-$\pi$- and $p$-$K$-scattering very well by using extension parameters which are quite similar to the electromagnetic radii of the different hadrons. By varying the extension parameter of the proton slightly with $s$ we obtained the right energy dependency of $\si^{\rm tot}_{p\bar{p}}$ and $B_{p\bar{p}}$ up to highest energies.
\section{The odderon exchange and the suppression through a diquark-structure}
\begin{figure}[ht]
\begin{minipage}{5.5cm}
\epsfxsize5.5cm
\epsffile{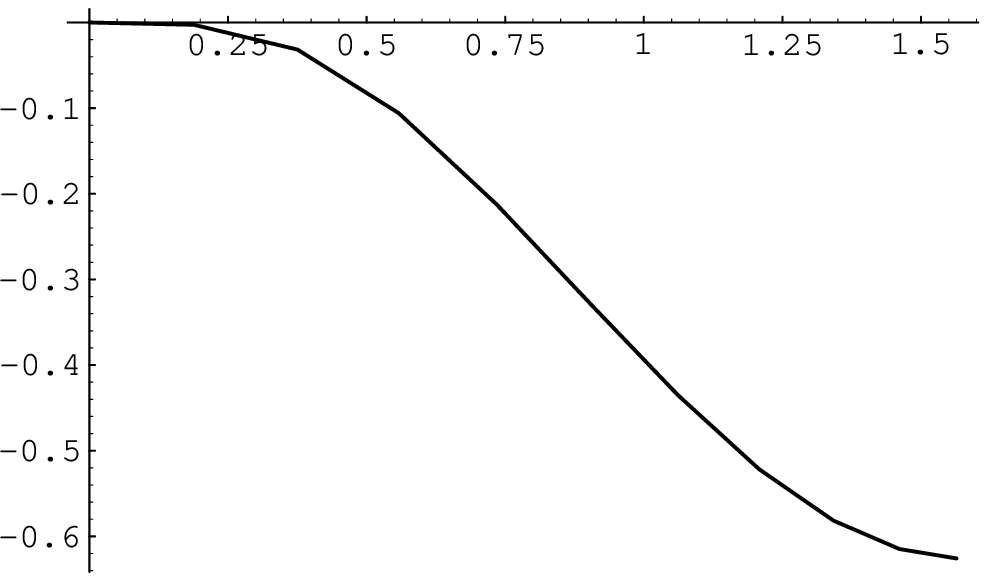}
\end{minipage}
\hfill
\begin{minipage}{5.5cm}
\epsfxsize5.5cm
\epsffile{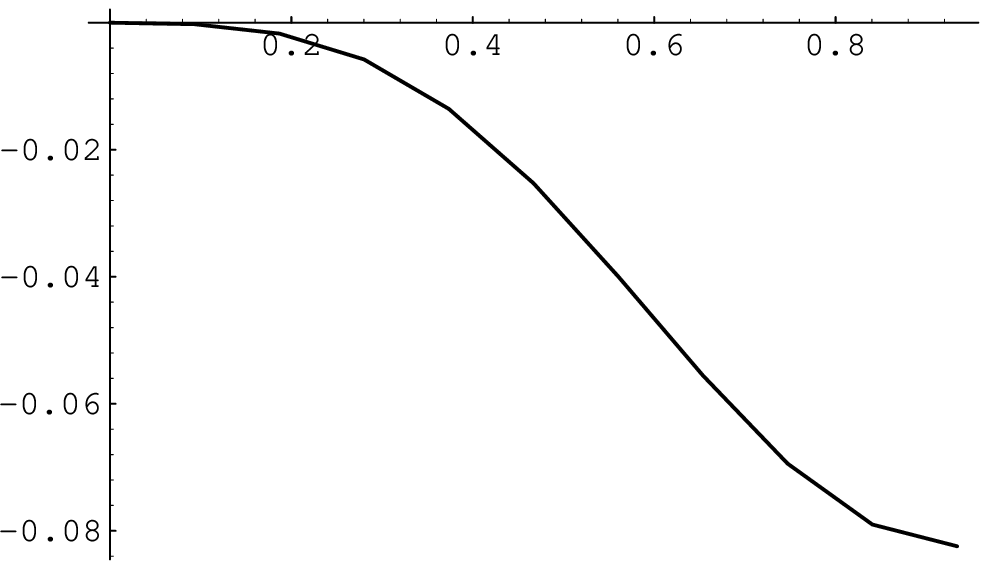}
\end{minipage}
\centering
\caption{$\De \rh$ at UA4/2 energy for proton-(anti)proton scattering as a function of the quark distances $r_\perp$ and $r_\|$ (see \fig{ort}). Left: star-like geometry ($S_p$=0.90 fm). Right: linear geometry ($S_p$=0.93 fm). Note the different scales.}
\label{rhoplots}
\unitlength1cm
\begin{picture}(0,0)
\put(1.2,1.9){$\De \rh$}
\put(-1.6,4.3){$r_\perp[\fm]$}
\put(-5.3,1.9){$\De \rh$}
\put(5,4.3){$r_\|[\fm]$}
\end{picture}
\end{figure}

In this section we calculate the leading contribution to the scattering amplitude which is odd under charge parity. From \eq{Jmeson} it is easy to see that for a meson or equivalently a baryon in the diquark picture the contribution of the \mbox{$C$=$P$=--1} exchange is appreciable for a given Wilson-loop. Constructing the hadrons by averaging the loops with wave functions ,however, cancels these contributions. So we use now the two $qqq$-configurations of the baryon mentioned before (see \fig{ort}). By expanding the exponentials of the Wilson-loops in \eq{Jbaryon} up to third order, we find\ct{Rueter:1996} that the leading contribution with odd charge parity to the scattering amplitude is real. We still get the same total cross section for hadron-hadron- and hadron-antihadron-scattering but in contrast to the leading order we find a non-vanishing contribution to the rho-parameter:
\[
\De \rh(s) =\frac{{\rm Re}\left[ T^{\bar p p}(s,0)\right]}{{\rm Im}\left[  T^{\bar p p}(s,0)\right]}-\frac{{\rm Re}\left[ T^{ p p}(s,0)\right]}{{\rm Im}\left[ T^{ p p}(s,0)\right]}=-2\frac{{\rm Re}\left[ T^{ p p}(s,0)\right]}{{\rm Im}\left[ T^{ p p}(s,0)\right]}^{C=-}.
\]
The result for the two different baryon configurations as a function of the diquark size at UA4/2 energy ($\sqrt{s}$=541 GeV) is shown in \fig{rhoplots}.
\\As can be seen from \fig{rhoplots} we obtain for the symmetric star-like geometry a result which is very similar to the value obtained in the Abelian non-perturbative gluon exchange model\cite{Donnachie:1991}. A clustering of two quarks to a diquark with an extension smaller or equal to $0.3$ fm yields on the other hand already a drastic suppression of $\De \rh$ to a value $|\De \rh | \le 0.02$ which is compatible with the analysis of experiments. There is plenty of other evidence for diquark clustering in baryons: The scaling violation in nucleon structure functions\ct{Anselmino:1993}, the strong attraction in the scalar diquark channel in the instanton vacuum\ct{Schafer:1994} and the $\De I=\frac{1}{2}$ enhancement in semi leptonic decays of baryons\ct{Dosch:1989}. Scaling violation in deep inelastic scattering is sensible to the form factors of the diquark, which is modeled by a pole fit with a pole mass of $\sqrt{3}$ to $\sqrt{10}$ GeV. This corresponds to diquark radii of 0.3 to 0.16 fm. These values are according to our model sufficiently small to give a suppression of $|\De \rh|$ to values below 0.02 even for the transversal extension.\\
Our calculation has been performed in a specific non-perturbative model. But since the limiting case of a vanishing diquark radius leads quite generally to an odderon cancelation as in the case of mesons (see \eq{3c} and the discussion of it) we think that the suppression of the $C$=$P$=--1 exchange is generally due to the structure of the nucleon and cannot be seen on the quark level.
\section*{Acknowledgment}
This work was done in collaboration with H.G.~Dosch. The author also thanks M.~Anselmino for the invitation to give this talk. Finally the author is grateful to the Deutsche Forschungsgemeinschaft and to the Bundesland Baden-W\"urttemberg for financial support.


\begin{thebibliography}{10}
\let\qq=\"

\bibitem{Rueter:1996}
M.~Rueter and H.G. Dosch, {\it Phys.Lett.\/} {\bf B380} (1996) 177.

\bibitem{Lukaszuk:1973}
L.~Lukaszuk and B.~Nicolescu, {\it Lett.Nuov.Cim.\/} {\bf 8} (1973) 405.

\bibitem{Bernard:1987}
{D.~Bernard et al.}, {\it Phys.Lett.\/} {\bf B198} (1987) 583, UA4
  Collaboration.

\bibitem{Augier:1993}
{C.~Augier et al.}, {\it Phys.Lett.\/} {\bf B316} (1993) 448, UA4/2
  Collaboration.

\bibitem{Bourrely:1984}
C.~Bourrely and A.~Martin, in {\it CERN-ECFA Workshop, Lausanne\/} (1984).

\bibitem{Donnachie:1984}
A.~Donnachie and P.V. Landshoff, {\it Nucl.Phys.\/} {\bf B244} (1984) 322.

\bibitem{Bourrely:1987}
C.~Bourrely, J.~Soffer and T.T. Wu, {\it Phys.Lett.\/} {\bf B196} (1987) 237.

\bibitem{Gauron:1988}
P.~Gauron, B.~Nicolescu and E.~Leader, {\it Nucl.Phys.\/} {\bf B299} (1988)
  640.

\bibitem{Kroll:1989}
P.~Kroll and W.~Schweiger, {\it Nucl.Phys.\/} {\bf A503} (1989) 865.

\bibitem{Covolan:1993}
R.J.M.~Covolan et~al., {\it Z.Phys.\/} {\bf C58} (1993) 109.

\bibitem{Donnachie:1992}
A.~Donnachie and P.V. Landshoff, {\it Phys.Lett.\/} {\bf B296} (1992) 227.

\bibitem{Landshoff:1987}
P.V. Landshoff and O.~Nachtmann, {\it Z.Phys.\/} {\bf C35} (1987) 405.

\bibitem{Donnachie:1991}
A.~Donnachie and P.V. Landshoff, {\it Nucl.Phys.\/} {\bf B348} (1991) 297.

\bibitem{Dosch:1994}
H.G. Dosch, E.~Ferreira and A.~Kr{\"a}mer, {\it Phys.Rev.\/} {\bf D50} (1994)
  1992.

\bibitem{Dosch:1987}
H.G. Dosch, {\it Phys.Lett.\/} {\bf B190} (1987) 177.

\bibitem{Dosch:1988}
H.G. Dosch and Y.A. Simonov, {\it Phys.Lett.\/} {\bf B205} (1988) 339.

\bibitem{Nachtmann:1991}
O.~Nachtmann, {\it Annals Phys.\/} {\bf 209} (1991) 436.

\bibitem{Rueter:1997}
M.~Rueter, {\it Quark-Confinement und diffraktive Hadron-Streuung im Modell des
  stochastischen Vakuums\/}, PhD-Thesis at the University of Heidelberg (1997).

\bibitem{Anselmino:1993}
M.~Anselmino et~al., {\it Rev.Mod.Phys.\/} {\bf 65} (1993) 1199.

\bibitem{Schafer:1994}
T.~Sch{\"a}fer, E.V. Shuryak and J.J.M. Verbaarschot, {\it Nucl.Phys.\/} {\bf
  B412} (1994) 143.

\bibitem{Dosch:1989}
H.G. Dosch, M.~Jamin and B.~Stech, {\it Z.Phys.\/} {\bf C42} (1989) 167.

\end{thebibliography}
\end{document}